\documentclass[prd,onecolumn,notitlepage,showpacs,preprintnumbers,amsmath,amssymb,nofootinbib,APS,10pt,superscriptaddress]{revtex4-1}

\usepackage{dcolumn}
\usepackage{bm}
\usepackage{xcolor}
\usepackage[utf8]{inputenc}
\usepackage[spanish,english]{babel}
\usepackage{amsmath,amssymb,amsfonts,latexsym,cancel}
\usepackage{graphicx}
\usepackage{color}
\usepackage{soul}
\usepackage{ulem}
\usepackage{hyperref}
\usepackage{amsmath}
\usepackage{slashed}
\usepackage{braket}

\allowdisplaybreaks

\newcommand{\be}{\begin{equation}}
\newcommand{\ee}{\end{equation}}
\newcommand{\bea}{\begin{eqnarray}}
\newcommand{\eea}{\end{eqnarray}}

\begin{document}

\title{{\bf 
Breaking of 
adiabatic invariance in the creation of particles by electromagnetic backgrounds}} 

\author{Pau Beltr\'an-Palau}\email{pau.beltran@uv.es}
\author{Antonio Ferreiro}\email{antonio.ferreiro@ific.uv.es}
\author{Jose Navarro-Salas}\email{jnavarro@ific.uv.es}
\author{Silvia Pla}\email{silvia.pla@uv.es}

\affiliation{Departamento de Fisica Teorica and IFIC, Centro Mixto Universidad de Valencia-CSIC. Facultad de Fisica, Universidad de Valencia, Burjassot-46100, Valencia, Spain.}

\begin{abstract}
Particles are spontaneously created from the vacuum by time-varying gravitational or electromagnetic backgrounds. It has been  proven that the particle number operator in an expanding universe is  an adiabatic invariant. 
In this paper we show that, in some special cases, 
the expected adiabatic invariance of the particle number fails  in presence of electromagnetic backgrounds. In order to do this, we consider as a prototype  a Sauter-type electric pulse. Furthermore, we also show a close relation between the breaking of the adiabatic invariance and the emergence of the axial anomaly.
\\

{\it Keywords:}  Particle creation, Schwinger effect,  adiabatic invariance, adiabatic regularization, axial anomaly.
 \end{abstract}

\pacs{04.62.+v, 11.10.Gh, 12.20.-m, 11.30.Rd}

\date{\today}
\maketitle
\section{Introduction}\label{Introduction}

The  understanding of particle creation phenomena in terms of  Bogolubov transformations was pioneered in the analysis of quantized fields in an isotropically expanding universe \cite{parker66, parker-toms, birrell-davies} (for a retrospective analysis see \cite{parker2012}). A fundamental issue  in the study of  particle creation in an expanding universe was the  adiabatic invariance of the number of created particles.  The particle number of a  quantized field, in the limit of an infinitely slow and smooth expansion of the universe, that is, an adiabatic expansion,
does not change with time \cite{parker2012}, even if the quantized field is massless. In other words, the density of created particles by the  cosmic expansion approaches zero when the Hubble rate $\dot a/a$ is each time  negligible even if the final amount of expansion $a(t_{final})/a(t_{initial})$ is large.   Hence, we say that the particle number is an adiabatic invariant. Moreover, pair production can also take place in time-varying electric \cite{Pittrich-Gies, ELI} or scalar backgrounds, and it can be regarded as a very important non-perturbative process in quantum field theory \cite{Schwinger51}. It is also fundamental to understand  the reheating epoch in cosmology \cite{reheating}, non-equilibrium processes induced by strong fields \cite{Mueller, Dunne}, and astrophysical phenomena \cite{R}.  \\

The main purpose of this work is to analyze the adiabatic invariance of the particle number observable in the presence of an electromagnetic background.  We find that for massive fields adiabatic invariance is, as expected, preserved. For slowly varying electromagnetic potentials  no quanta is being produced, even if the change in the electromagnetic potential 
over a long period is very large. However, in some cases and only for massless fields,  the particle number is not an adiabatic invariant. In other words, particles are still created in the adiabatic limit. We analyze the problem in detail in a two-dimensional scenario, for both scalar and Dirac fields. As a by-product of our analysis, we  point out a connection between the (anomalous) breaking of the adiabatic invariance of the particle number operator and the emergence of a quantum anomaly in the chiral symmetry.  We will show that the breaking of adiabatic invariance and its connection to the axial anomaly can be easily translated  to four dimensions.  \\

Conservation laws and symmetries  play a fundamental role in the understanding of a physical system. Anomalies are symmetries of a classical theory that  fail to survive upon quantization.  
This happens, typically, in field theory because of the need for regularization and renormalization of ultraviolet divergences. 
A very illustrative example occurs  in quantum electrodynamics  in the limit of massless Dirac fermions. The classical theory is invariant under chiral transformations and this implies the conservation of the axial current $j^\mu_5$.  However, this symmetry is broken in the quantum theory. The chiral anomaly opens the possibility of having processes violating the conservation of chirality. Nevertheless, all elementary processes of quantum electrodynamics, based on the perturbative expansion of the S-matrix, preserve chirality \cite{qftbook}. One has to resort to a non-perturbative phenomena, i.e., the  spontaneous pair production by electromagnetic fields,
to unveil conservation-law violation of chirality of massless fermions. The non-conservation of chirality seems to be directly related to  the breaking of  adiabaticity in the particle number observable.  \\

The paper is organized as follows. Section II is devoted to briefly illustrate  the problem within the conventional cosmological scenario, as described in  \cite{birrell-davies,parker2012}. In Section III we will analyze the case  of a two-dimensional complex scalar field coupled to an external electric pulse. The role of the mass is analyzed in detail and we will show explicitly that  adiabatic invariance of the particle number is broken for massless fields. In Section IV we generalize the result to Dirac fields, showing a connection between the breaking of adiabatic invariance and the emergence of the chiral anomaly. The next step is to extend our result to four dimensions. This will be done in Section V. We will find again that  adiabatic invariance requires a non-vanishing effective mass, as happens for  two-dimensional quantized fields coupled to an electric field.   However, a  zero effective mass can only be  achieved for Dirac (not for scalar) fields coupled to both electric and magnetic fields. The breaking of adiabatic invariance also emerges in parallel to the  emergence of the chiral anomaly. In section VI we summarize the main conclusions.

\section{A brief orientation: Adiabatic invariance 
in the expanding universe}\label{Sec1}

\begin{figure}[htbp]
\begin{center}
\begin{tabular}{c}
\includegraphics[width=70mm]{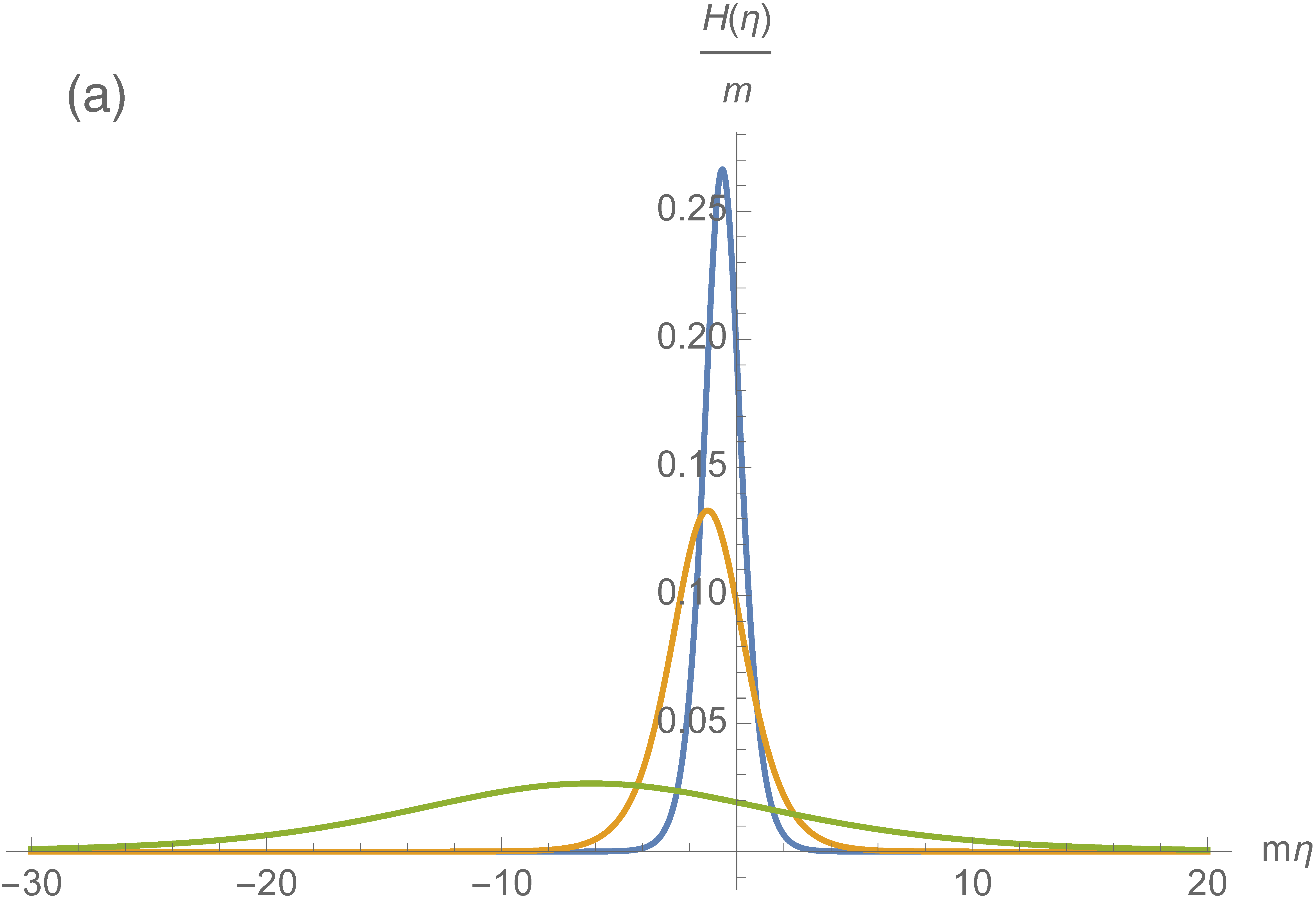}
\includegraphics[width=86mm]{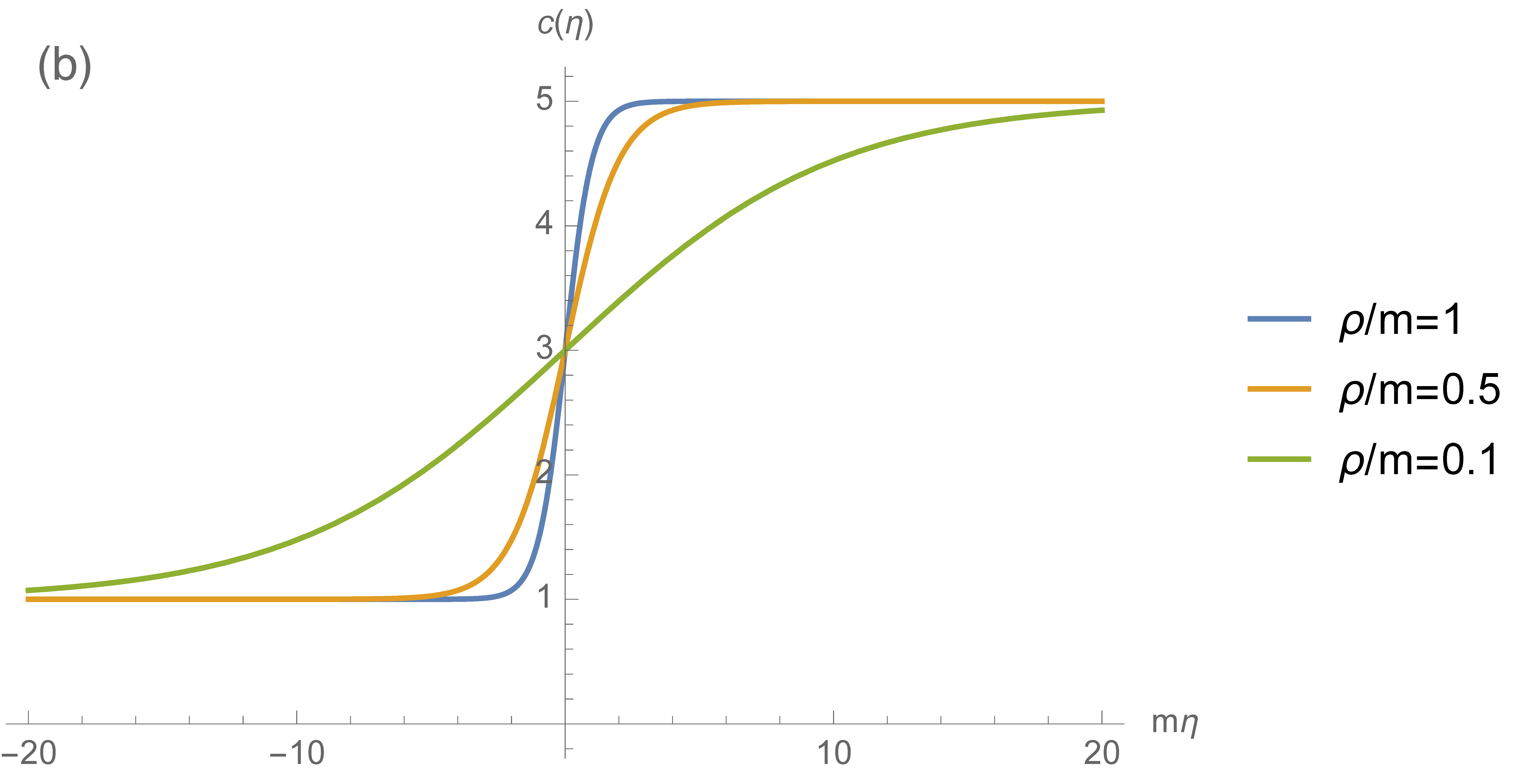}
\end{tabular}
\end{center}
\caption{\small{Conformal scale factor for $B=2$. Figure (a) shows the Hubble rate $H/m$ for different values of the dimensionless  ``slowness'' parameter $\rho/m$.  Figure (b) shows the dependence of the conformal scale factor $C(\eta)$ on $\rho/m$. The adiabatic limit corresponds to $\rho \to 0$. Note that the area defined by the curves  $H(\eta) $ in Figure (a) does not depend on  $\rho/m$.  }}
\label{conformal1}
\end{figure}

The  adiabatic invariance of the particle number operator in an expanding universe can be easily illustrated with the simple example (borrowed from \cite{birrell-davies}) of a scalar field with mass $m$ in the presence of a two-dimensional bounded expanding universe. This example, although well-known, will serve to better clarify the main idea of the next sections. Consider the following metric:
\be \label{metric1}ds^2 = dt^2 - a^2(t)dx^2 = C(\eta) (d\eta^2 -dx^2) \ , \ee
where $d\eta= a^{-1}(t) dt$ and the conformal scale factor is given by the function $ \label{C} C(\eta) = 1+ B(1+\tanh \rho \eta)$,
with $B$ a positive constant.
This represents a smooth  expansion  bounded by asymptotically static and flat spacetime regions.  
The expansion factor has smoothly shifted from $a_{in}\equiv a(-\infty)=1$ to $a_{out}\equiv a(+\infty)=\sqrt{1+2B}$.  In Fig. \ref{conformal1} it is shown the behavior of the conformal scale factor $C(\eta)$ as well as the Hubble rate $H(\eta)=\frac{C'(\eta)}{2C^{3/2}}$ for different values of the adiabatic parameter $\rho$ in terms of dimensionless variables. 

The equation for the modes of the scalar field in the background metric (\ref{metric1}) is given by
\be
\frac{d^2}{d\eta^2}h_{k}(\eta)+\left(m^2C(\eta)+k^2\right)h_{k}(\eta)=0  \ , \label{equhkgrav}
\ee
In the remote past the normalized modes  are assumed to behave as the positive frequency modes in Minkowski space
\be \frac{1}{\sqrt{2(2\pi)\omega_{in}}} e^{ikx} e^{-i\omega_{in} t}\ , \ee
with $\omega_{in} = \sqrt{k^2 + m^2}$. 
As time evolves these modes behave, in the remote future, as a mixture of positive and negative frequency modes of the form
\be \frac{\alpha_k}{\sqrt{2(2\pi)\omega_{out}}} e^{ikx} e^{-i\omega_{out} t} + \frac{\beta_k}{\sqrt{2(2\pi)\omega_{out}}} e^{ikx} e^{+i\omega_{out} t} \ , \ee
with  $\omega_{out} = \sqrt{(\frac{k}{a_{out}})^2 + m^2}$. $\alpha_k$ and $\beta_k$ are the so-called Bogolubov coefficients.  The annihilation operators for physical particles at late times $a_k$ are related to the annihilation and creation operators at early times ($A_k$ and $A_k^{\dagger}$) by the relations
\be a_k = \alpha_k A_k+ \beta^*_k A^\dagger_{-k} \ . \ee
The average density number of created particles $ n_k$, with momentum $k$,  is given by
\be n_k =  |\beta_k|^2 =  \frac{\sinh^2(\pi \frac{\omega_- }{\rho})}{\sinh (\pi \frac{\omega_{in}}{\rho}) \sinh (\pi \frac{a_{out} \omega_{out}}{\rho})}\label{betagravity}\ , \ee
where  $\omega_{-}=\frac12 (a_{out}\omega_{out} - \omega_{in})$. It is very easy to check that in the adiabatic limit, that is, for an extremely slow expansion $\rho \to 0$, the density number of created particles goes to  $n_k \sim  e^{-2 \pi \omega_{in}/\rho} \to 0$. This shows the fact that the particle number is  an adiabatic invariant. 
This  behavior of the particle number observable is generic,  and it can be extended to  isotropically expanding universes in four dimensions, irrespective of the value of the mass \cite{parker66, parker2012}.\\

 Furthermore, one can reinforce this idea by looking at a gravitational collapse producing a black hole.  
An adiabatic collapse can be thought as the (physically inaccessible) limit of a collapse approaching to a black hole with very large mass $M \to \infty$ (and zero surface gravity) in an infinite amount of advanced time \cite{Israel}. It is well-known that the late-time particle creation of a gravitational collapse  is encapsulated by the surface gravity parameter. The produced radiation is thermal \cite{Hawking}-\cite{parker-toms, birrell-davies, FabbriNavarro}, with a temperature proportional to the surface gravity. In the adiabatic limit the production of scalar particles is expected to vanish, in agreement with Hawking's result.

 \section{Breaking of adiabatic invariance in scalar pair production by electric fields in two dimensions} 

\begin{figure}[htbp]
\begin{center}
\begin{tabular}{c}
\includegraphics[width=70mm]{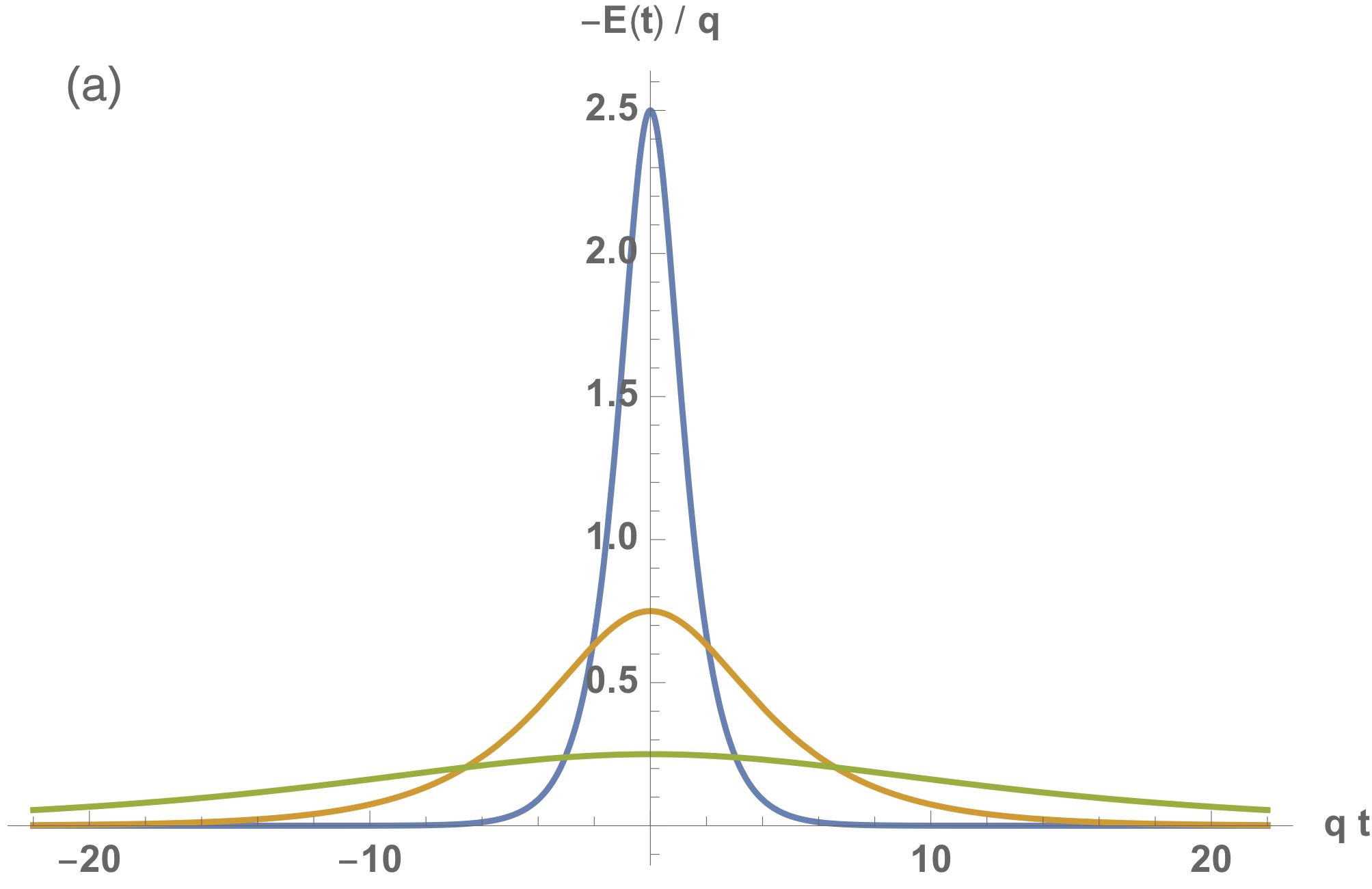}
\includegraphics[width=86mm]{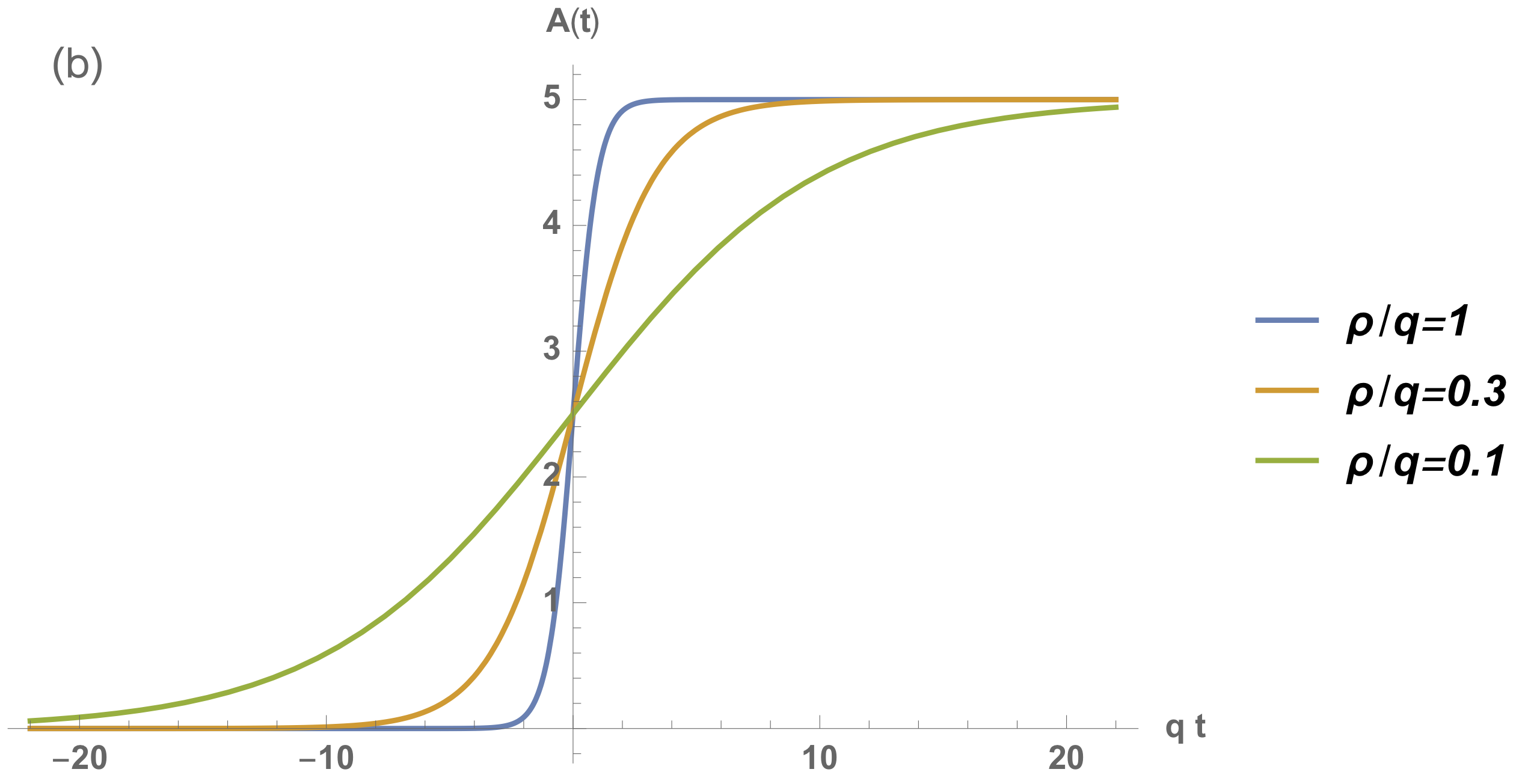}
\end{tabular}
\end{center}
\caption{\small{Sauter-type electric pulse for $A_0=5$. Figure (a) shows the electric field $E/q$ for different values of $\rho/q$.  Figure (b) shows the dependence of the  potential $A(t)$ on the dimensionless ``slowness'' parameter $\rho/q$. The adiabatic limit corresponds to $\rho \to 0$. Note that the area defined by the curves  $-E(t)/q $ in Figure (a) is $A_0$, irrespective of the value of $\rho/q$.  }}
\label{electricfield}
\end{figure}

We will now analyze the same question for the phenomena of particle creation in electric fields.  We will consider a classical and homogeneous electric field $E(t)$ interacting with a quantum, two-dimensional charged scalar field $\phi$ obeying the field equation
\be (D_\mu D^\mu + m^2) \phi =0 \label{motionscalar2}\ , \ee
where $D_\mu \phi = (\partial_\mu +iq A_\mu) \phi$. We can expand the field in Fourier modes as
\be \phi(t,x)= \frac{1}{\sqrt{2(2 \pi)}}\int d k [A_{k}e^{i k x}h_{k}(t)+B_{k}^{\dagger}e^{-i k x}h^*_{- k}(t) ] \ , \label{phisolution} \ee
where
 $A_{k}^{\dagger}, B_{k}^{\dagger}$ and $A_{k}, B_{k}$ are the usual creation and annihilation operators. The mode functions $h_{ k}(t)$ must obey  the Wronskian consistency condition
 \be \label{Wc} h_{ k}\dot h_{k}^* - h_{ k}^*\dot h_{ k} = 2i \  \ee 
 to ensure the usual commutation relations.
 Substituting \eqref{phisolution} into \eqref{motionscalar2} we get the equation
\be
\ddot{h}_{k}(t)+\left(m^2+(k-q A(t))^2\right)h_{k}(t)=0  \ , \label{equhk} \ee
   where we have chosen an homogeneous time dependent potential $A_{\mu}=(0, -A(t))$ in the appropriate  gauge. In order to study the adiabatic limit for the electric pair production, in a  way similar to the gravitational case explained above, we need to consider a {\it bounded  potential} $A(t)$. At an heuristic level,  $A(t)$ will play a somewhat similar role to the conformal factor $C(\eta)$  for the expanding spacetime. Note by comparing (\ref{equhkgrav}) and (\ref{equhk}) that the time dependence of the mode equation is encoded in  $C(\eta)$ for the gravitational example and, analogously,  it is  in  $A(t)$ in the electric case (see, for instance, \cite{Pittrich-Gies} for a general discussion).  We choose for convenience a Sauter-type electric pulse \cite{Sauter} of the form
   \be E(t) =  -\frac{\rho A_0}{2} \cosh^{-2} (\rho t) \label{Erho} \ , \ee
which can be described by the potential ($E(t) = -\dot A(t)$)
    \be A(t)=\frac12 A_0\left(\tanh(\rho t)+1\right) \label{potentiala} \ . \ee
 This potential is bounded both at early and late times, as shown explicitly in Fig.\ref{electricfield}(b). Note that $\rho$ plays the role of a slowness parameter. It is very illustrative to compare Fig.\ref{electricfield} with Fig.\ref{conformal1}\ . 
 
We have chosen the above Sauter-type pulse \cite{Sauter} for convenience. Note that this potential is bounded both at early and late times (see Fig.\ref{electricfield}). Note also that for all the figures we work with dimensionless variables.
 The adiabatic limit is an extremely slow evolution of the potential, obtained when $\rho \to 0$. We have to remark that the adiabatic limit is not the limit of a vanishing electric field. If the electric field had support in a bounded period of time, there would not be production of particles when $E (t) \to 0$. But the adiabatic limit is a more subtle limit, in which the electric field  varies very  slowly. Although $E \to 0$ when $\rho \to 0$, the  width of the pulse is also  very large   maintaining  constant and non-vanishing the integral
 \be \int_{-\infty}^{+\infty} E_{\rho_1}(t) dt = \int_{-\infty}^{+\infty} E_{\rho_2}(t) dt = \ \ constant = -qA_0\ . \ee 
 
To clarify things  we remark that a different scenario is given by the alternative choice $E(t) =  -E_0 \cosh^{-2} (\rho t)$, with $E_0$ a constant value, independent of $\rho$. The limit $\rho \to 0$ corresponds then to  a constant electric field, with an {\it unbounded potential} $A(t)$. This produces, as expected,  the Schwinger-type  rate of pair creation by a constant electric field \cite{FN}. In this paper we focus our analysis in the  adiabatic limit $\rho \to 0$ in (\ref{Erho}) and (\ref{potentiala}), as it  produces a bounded potential and a completely analogous situation to that considered in the cosmological scenario.  As we will show later on, in this case, it is indeed possible to produce particles by the electric field if extra conditions are met (i.e., a massless field, or the presence of magnetic fields in the four-dimensional case with fermions). \\

 Inserting the potential \eqref{potentiala} in \eqref{equhk} we obtain the physical solution in terms of the usual hypergeometric functions
 \bea
 h_k(t)=\frac{1}{\sqrt{\omega_{in}}}e^{-i \omega_{in} t}(1+e^{2 \rho t})^{(\frac12-i \frac{\kappa}{\rho})}F\left(\frac12 - i\frac{\omega_{+}+\kappa}{\rho}, \frac12 + i\frac{\omega_{-}-\kappa}{\rho},1-i\frac{\omega_{in}}{\rho};-e^{2\rho t}\right)\label{modeh}
 \eea
 where $\kappa=\frac12\sqrt{(qA_0)^2-\rho^2}$, $\omega_{in}=\sqrt{k^2 + m^2}$, $\omega_{out}=\sqrt{(k-qA_0)^2 + m^2}$  and $\omega_{\pm}=\frac12 (\omega_{out} \pm \omega_{in})$. We have fixed this solution by demanding that at early times  the modes behave as the Minkowskian modes for a free scalar field 
 \bea
 h_k(t)\sim\frac{1}{\sqrt{\omega_{in}}}e^{-i \omega_{in} t} \ .
 \eea
 At late times the modes behave as 
\be h_k(t)  \sim \frac{\alpha_k}{\sqrt{\omega_{out}}}  e^{-i\omega_{out} t} + \frac{\beta_k}{\sqrt{\omega_{out}}}  e^{+i\omega_{out} t} \label{latelimit}\ , \ee
where $\alpha_k$ and $\beta_k$ are the  Bogoliubov coefficients. They  satisfy the relation $|\alpha_k|^2-|\beta_k|^2=1$ according to the normalization condition \eqref{Wc}.  These coefficients serve to relate the early time creation and annihilation operators ${A_k,B_k}$, defining the initial Fock space, with the late time operators ${a_k,b_k}$
\bea
&&a_k=\alpha_k A_k+\beta_{k}^{*}B_{-k}^{\dagger}\\
&&b_k=\alpha_{-k}B_k+\beta_{-k}^{*} A_{-k}^{\dagger} \ .
\eea
Therefore, we can define the number operator as
\bea \label{Ns}
\langle N \rangle=\frac{1}{2 \pi}\int_{-\infty}^{\infty}dk~ N_k=\frac{1}{2 \pi}\int_{-\infty}^{\infty}dk\left(|\beta_{k}|^2+|\beta_{-k}|^2\right) \ , 
\eea
where $N_k=n_k+\bar n_k= \bra{0}a_k^{\dagger}a_k\ket{0}+\bra{0}b_k^{\dagger}b_k\ket{0}$ is the number density of quanta (i.e., $n_k= |\beta_k|^2$ particles  and  $\bar n_k=|\beta_{-k}|^2$ antiparticles).
Taking the late time limit $t \to \infty$ in \eqref{modeh} and matching with \eqref{latelimit} we obtain
\bea
\alpha_k=\sqrt{\frac{\omega_{out}}{\omega_{in}}}\frac{\Gamma(1-i\frac{\omega_{in}}{\rho})\Gamma(-i \frac{\omega_{out}}{\rho})}{\Gamma(\frac12 - i\frac{\omega_{+}+ \kappa}{\rho})\Gamma(\frac12 -i\frac{\omega_{+}- \kappa}{\rho})}\\
\beta_k=\sqrt{\frac{\omega_{out}}{\omega_{in}}}\frac{\Gamma(1-i\frac{\omega_{in}}{\rho})\Gamma(i \frac{\omega_{out}}{\rho})}{\Gamma(\frac12 + i\frac{\omega_{-}+ \kappa}{\rho})\Gamma(\frac12 +i\frac{\omega_{-}- \kappa}{\rho})}
\eea
where we have used the usual properties of the hypergeometric function \cite{mathbook}. Finally we get
 \bea |\beta_{k}|^2=\frac{\cosh{(2\pi \frac{\omega_{-}}{\rho})}+\cosh{(2\pi \frac{\kappa}{\rho})}}{2\sinh{(\pi \frac{\omega_{in}}{\rho})}\sinh{(\pi\frac{\omega_{out}}{\rho})}} \label{betaqed}\ .\eea

\begin{figure}[htbp]
\begin{center}
\begin{tabular}{c}
\includegraphics[width=90mm]{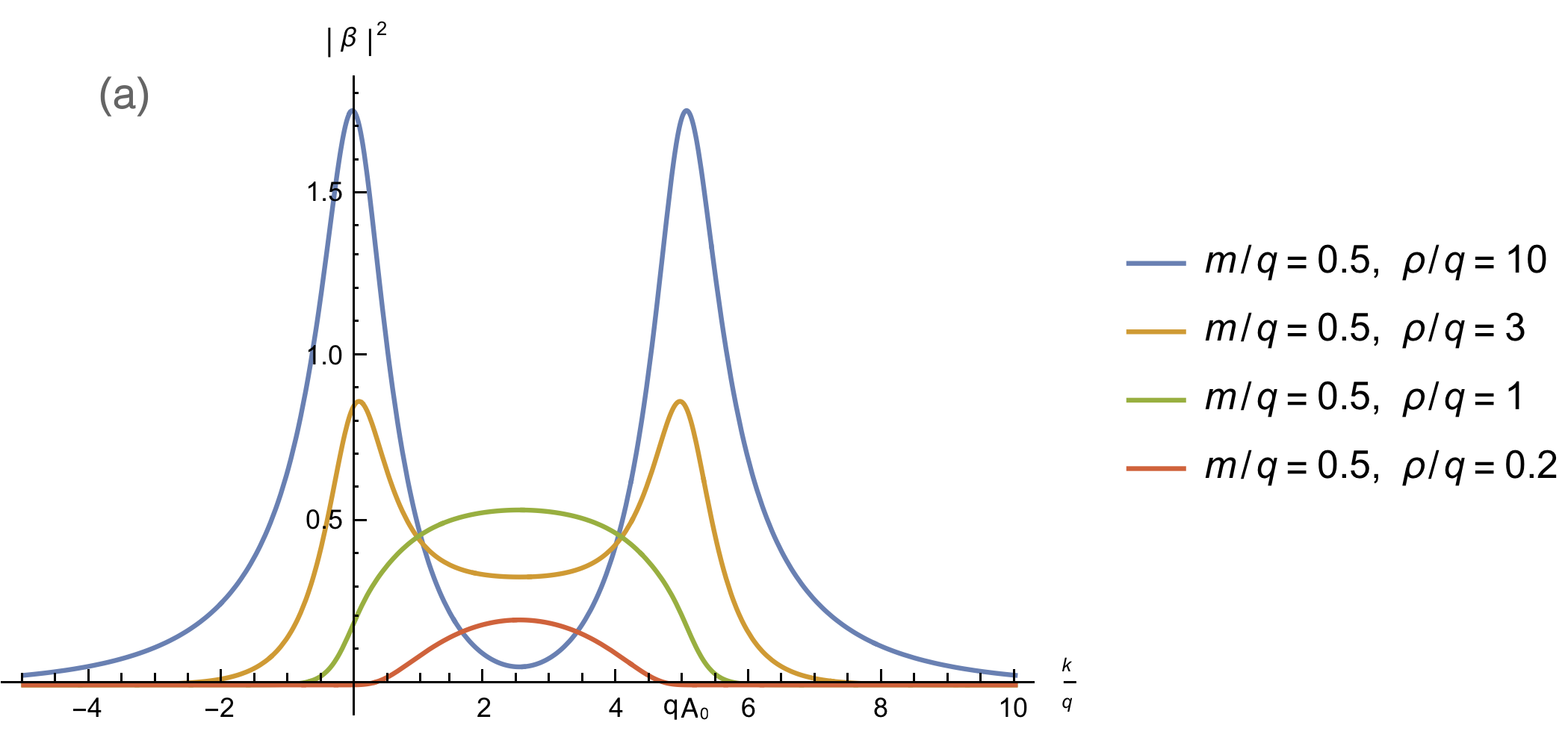}
\includegraphics[width=90mm]{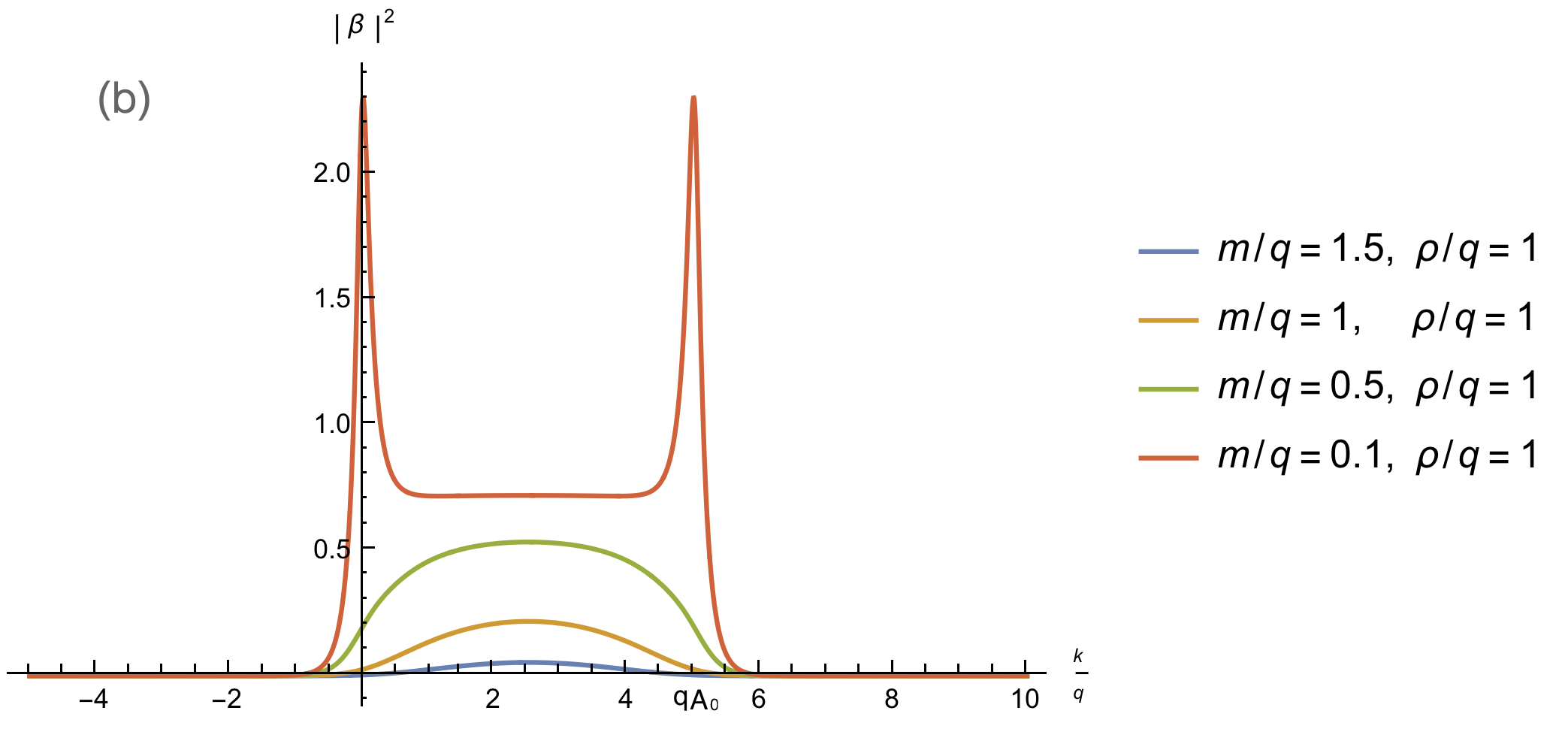}
\end{tabular}
\end{center}
\caption{\small{Momentum distribution of the created scalar particles with positive charge at late times ($|\beta_k|^2$) by an electric pulse with $A_0=5$ and different values of $m/q$ and $\rho/q$. In figure (a) the mass is fixed, while in (b) the dimensionless parameter of adiabaticity $\rho/q$ is fixed. }}
\label{betascalar}
\end{figure}

Fig. \ref{betascalar} shows a representation of this expression for different values of $m$ and $\rho$, which can be interpreted as the momentum distribution of the created particles (the spectra of antiparticles  would be obtained by making the shift $k\to-k$). We easily observe that $|\beta_k|^2$ decreases as $\rho \to 0$, for fixed $m\neq 0$. In the same way, the particle density also decreases for large $m$ with $\rho$ fixed. Note in passing  that for a  sudden  electric pulse ($\rho>>0$) the momentum distribution of the particles is concentrated in the characteristic values $k=0$ and $k=qA_0$.

\begin{figure}[htbp]
\begin{center}
\includegraphics[width=90mm]{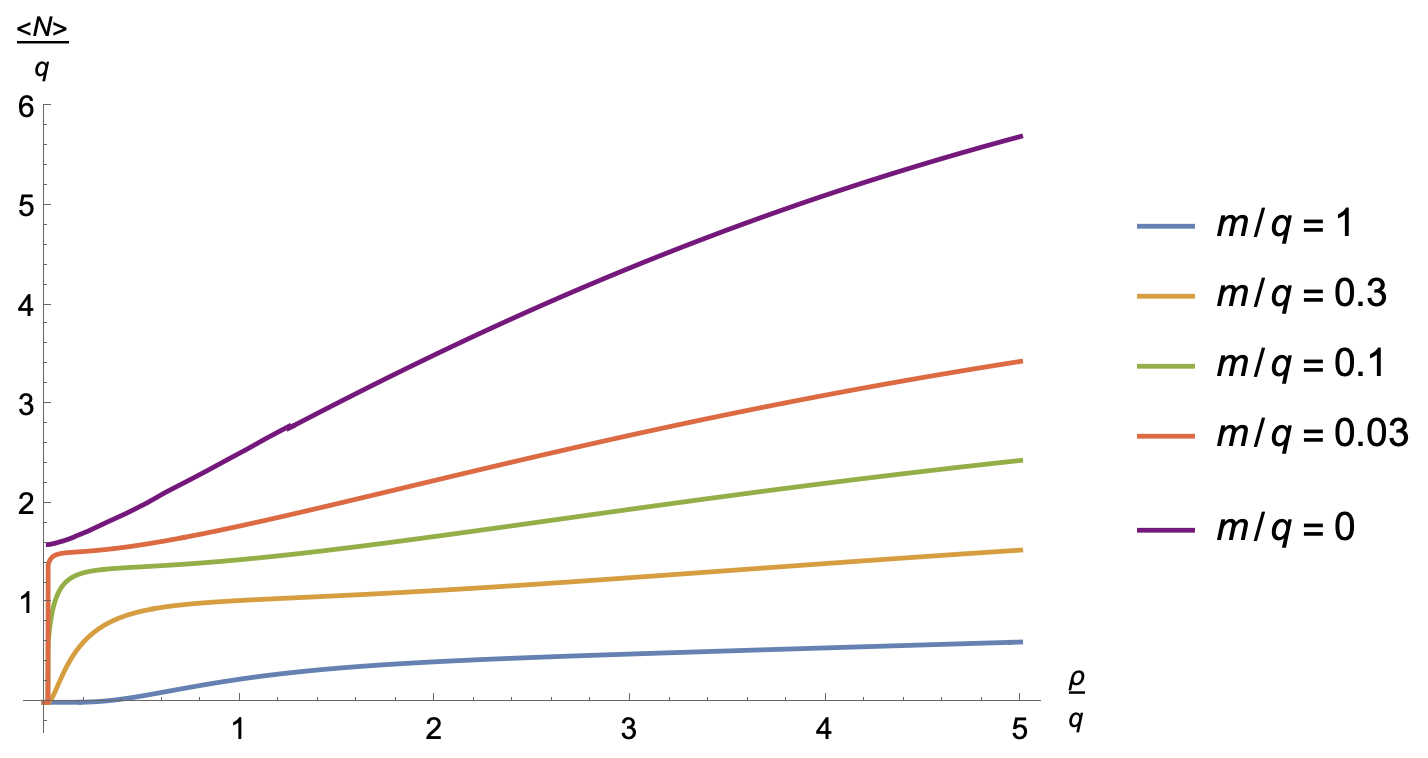}
\end{center}
\caption{\small{Number of late-time created scalar particles  as a function  of the dimensionless adiabaticity parameter  $\rho/q$, for $A_0=5$ and for different values of the mass. }}
\label{Nscalar}
\end{figure}

To see whether $|\beta_k|^2$   vanishes in the adiabatic limit we analyze in detail the behavior $\rho \to 0$ on \eqref{betaqed}. We get
\be
|\beta_k|^2  \sim e^{-2 \pi \omega_{in}/\rho}+e^{-2 \pi \omega_{out}/\rho} +e^{-\frac{\pi}{\rho} \delta} \ , \label{betalimit}
\ee
where $\delta=2(\omega_+-\kappa)$ and $\kappa \to \frac{|q A_0|}{2}$. Since $\omega_{in},\omega_{out}>0$, the first two terms vanish as $\rho \to 0$. For $m\neq0$, the function $\delta(k)$ has a minimum at  $k=\frac{qA_0}{2}$, with a value $\delta_{min}=\sqrt{(qA_0)^2+4m^2}-|qA_0|>0$. It means that $\delta>0$,  and hence $|\beta_k|^2 \to 0$ when $\rho \to 0$, as in the case of a gravitational field. According with that, in Fig. \ref{betascalar}(a) one can realize how the number of particles decreases with the adiabatic parameter $\rho$, vanishing in the limit $\rho\to0$. However, for $m=0$ this  is no longer valid  since $\delta=0$ for $k\in (0,qA_0)$, and hence $ |\beta_k|^2 \to 1$, meaning that particles are being produced even in the adiabatic limit. 
In short, we have obtained, when $\rho \to 0$,
\bea
N_k \to \left\{ \begin{array}{cc}
   0 & \text{ for } m\neq0 \ \ \text{or} \ m=0  \ \ \text{and} \  k\not\in(-qA_0,qA_0) \\
 1 & \text{ for } m=0 \text{ and } k\in(-qA_0,qA_0) \ \ . \\
 \end{array} \right. 
\eea 
In order to visualize this behavior, we represent in Fig. \ref{Nscalar} the dependence of the total density of created particles $\langle N\rangle$ (given by \eqref{Ns}) on the parameter $\rho$. One can see how in the adiabatic limit the density of quanta tends to vanish, except in the case $m=0$, for which it tends to a non-zero value. This value is given by

\bea
\langle N \rangle=\frac{1}{2\pi}\int_{-|q A_0|}^{|qA_0|} dk~ N_k=\frac{|qA_0|}{\pi} \label{particleprod}.
\eea
This implies that the particle number is not an adiabatic invariant for the massless case. Furthermore, as we will see in the next section, the above result for the density number of created particles in the adiabatic limit coincides exactly with the analogous result for massless Dirac particles.\\

For completeness we will  study now  the vacuum expectation values of the electric current and the energy density induced by the underlying particle creation process. This will also serve to test the adiabatic invariance, or the breaking of it, in terms of the current and the energy density. 

\subsection{Electric current}\label{EcurrentEsc}

For a two-dimensional charged scalar field, the electric current  is given by $j^{\mu}=iq\left[\phi^{\dagger}D^{\mu}\phi-(D^{\mu}\phi)^{\dagger}\phi \right] $. The vacuum expectation of this observable is UV-divergent and has to be renormalized. In the context of an homogeneous and time dependent background it is very convenient to use the adiabatic regularization/renormalization method described in \cite{FNP, FN}. After performing the appropriated subtractions, one obtains
\bea
&&\langle j^{x}\rangle_{ren}=q \int_{-\infty}^{\infty} \frac{dk}{2 \pi } \left [\left(k-q A\right) |h_k|^2-\frac{k}{\omega}+\frac{m^2 q A}{\omega^3} \right ] \label{renj} \ ,
\eea
where $h_k(t)$ are the mode functions of the scalar field satisfying the equation of motion \eqref{equhk} and $\omega=\sqrt{k^2+m^2}$. For more details on the original adiabatic method for scalar fields see \cite{parker-fulling}. \\

Let us  focus on the late-time behavior of the  electric current, for which we can relate \eqref{renj} to the Bogoliubov coefficients computed in the last section. We restrict again the analysis to an  electric-pulse configuration \eqref{potentiala} with bounded asymptotic states. Introducing \eqref{latelimit} in \eqref{renj} we have
\bea
\langle j^{x}\rangle_{ren}&\sim& q\int_{-\infty}^{\infty} \frac{dk}{2 \pi } \left [\frac{k-q A_0 }{\omega_{out}}\left(2|\beta_k|^2+2\mathrm{Re}(\alpha_k\beta_k^*e^{-i2\omega_{out} t}) \right)+\frac{k-q A_0 }{\omega_{out}}-\frac{k}{\omega}+\frac{m^2 q A_0}{\omega^3} \right]
 \label{jbeta} \ .
\eea
It is easy to see that the terms which are independent of the Bogoliubov coefficients do not contribute to the electric current. One can  derive this result by realizing  that the first two terms correspond to linearly divergent integrals, differing by a constant shift  
\be \label{ldi}q\int_{-\infty}^{\infty} \frac{dk}{2 \pi } \left [\frac{k-qA_0}{\sqrt{(k-qA_0)^2+m^2}}-\frac{k}{\sqrt{k^2 + m^2}}  \right ] =  -\frac{q^2A_0}{ \pi }  \ , \ee
while the last term in (\ref{jbeta}) is a finite integral, which cancels with (\ref{ldi}).
The second term in \eqref{jbeta} depends on time and produces oscillations of the form $\cos{(2\omega_{out}t)}$. In the limit $t\to\infty$ the Riemann-Lebesgue lemma ensures that the integral in $dk$ of this term vanishes. With the above considerations, and using the symmetry properties of $|\beta_k|^2$  (reflection under $k \to -(k-qA_0)$), one can rewrite the expression of the electric current as follows:
\bea \langle j^{x}\rangle_{ren}&\sim&-q\int_{-\infty}^{\infty} \frac{dk}{ 2\pi }  \frac{k }{\omega} (|\beta_k|^2-|\beta_{-k}|^2) \label{jreninf}\ .\eea
This equation shows explicitly the close relation between the density of created quanta and the electric current. The first term accounts for particles and the second one for antiparticles. In the adiabatic limit, and for massive particles, the renormalized electric current also vanishes since $|\beta_k|^2 \to 0$. 
However, the last result changes completely if  $m=0$. As we have shown, in the adiabatic limit
$|\beta_k|^2 \to 1$ for $k\in(0,qA_0)$.  Therefore, the current at late times  for massless particles in the adiabatic limit is given by
\bea
\langle j^{x}\rangle_{ren} \sim  -\frac{q^2 A_0}{\pi }\label{jbetam0} \ .
\eea
As expected, a non vanishing particle number $\langle N \rangle$, even in the adiabatic limit, induces  an electric current different from zero.

\subsection{Energy density}
The renormalized vacuum expectation value of the energy density of a two-dimensional scalar field interacting with an electric field is given by
\bea
\langle T_{00} \rangle_{ren} =\int_{-\infty}^{\infty}  \frac{dk}{4\pi} \left[|\dot{h}_k|^2+ \left(m^2+(k-q A)^2 \right)|h_k|^2-2 \omega +\frac{2 k q A}{\omega}-\frac{m^2 q^2 A^2}{\omega^3}\right]\label{engsc} \ ,
\eea
where $h_k(t)$ are again the mode functions of the scalar field and the three last terms account for the adiabatic subtractions required by renormalization \cite{FNP}.  As for the electric current, we will we focus on the late time behavior. Plugging \eqref{latelimit} in \eqref{engsc} and using the asymptotic expansion for the functions $\dot{h}(t)$  
\be \dot{h}_k(t)  \sim -i\sqrt{\omega_{out}} \alpha_k e^{-i\omega_{out} t} +i \sqrt{\omega_{out}} \beta_k e^{+i\omega_{out} t} \label{latelimithdot}\ , \ee
we finally obtain
\bea \label{rhobeta}
\langle T_{00}\rangle_{ren}&\sim&  \int_{-\infty}^{\infty}  \frac{dk}{4\pi} \left[4\omega_{out}|\beta_k|^2+2\omega_{out}-2\omega +\frac{2 k q A_0}{\omega}-\frac{m^2 q^2 A_0^2}{\omega^3}\right].
\eea

Using the same arguments as in section \ref{EcurrentEsc},  it is easy to see that the only term contributing to the energy density is the one proportional to $|\beta_k|^2$.  
After some simplifications, we get the relation between the energy density and the particle number
\bea \label{rhobetaSim}
\langle T_{00}\rangle_{ren} \sim\int_{-\infty}^{\infty} \frac{dk}{2\pi}\, \omega \,N_k \ ,
\eea
where $N_k=|\beta_{-k}|^2+|\beta_k|^2$. 
In the adiabatic limit, we get $|\beta_k|^2 \rightarrow 0$, and therefore $\langle T_{00}\rangle_{ren} \to 0$. Nevertheless, for $m=0$ there is indeed creation of energy. As we said, the adiabatic limit for the massless case gives us a non vanishing $|\beta_k|^2$ for $k\in(0,qA_0)$. In this region, $|\beta_k|^2=1$, and therefore the created energy density is

\bea
\langle T_{00}\rangle_{ren} \sim \frac{q^2 A_0^2}{2\pi }\label{Tbetam0} \ .
\eea

\section{Breaking of adiabatic invariance in fermionic pair production by electric fields in two dimensions
}

\begin{figure}[htbp]
\begin{center}
\begin{tabular}{c}
\includegraphics[width=90mm]{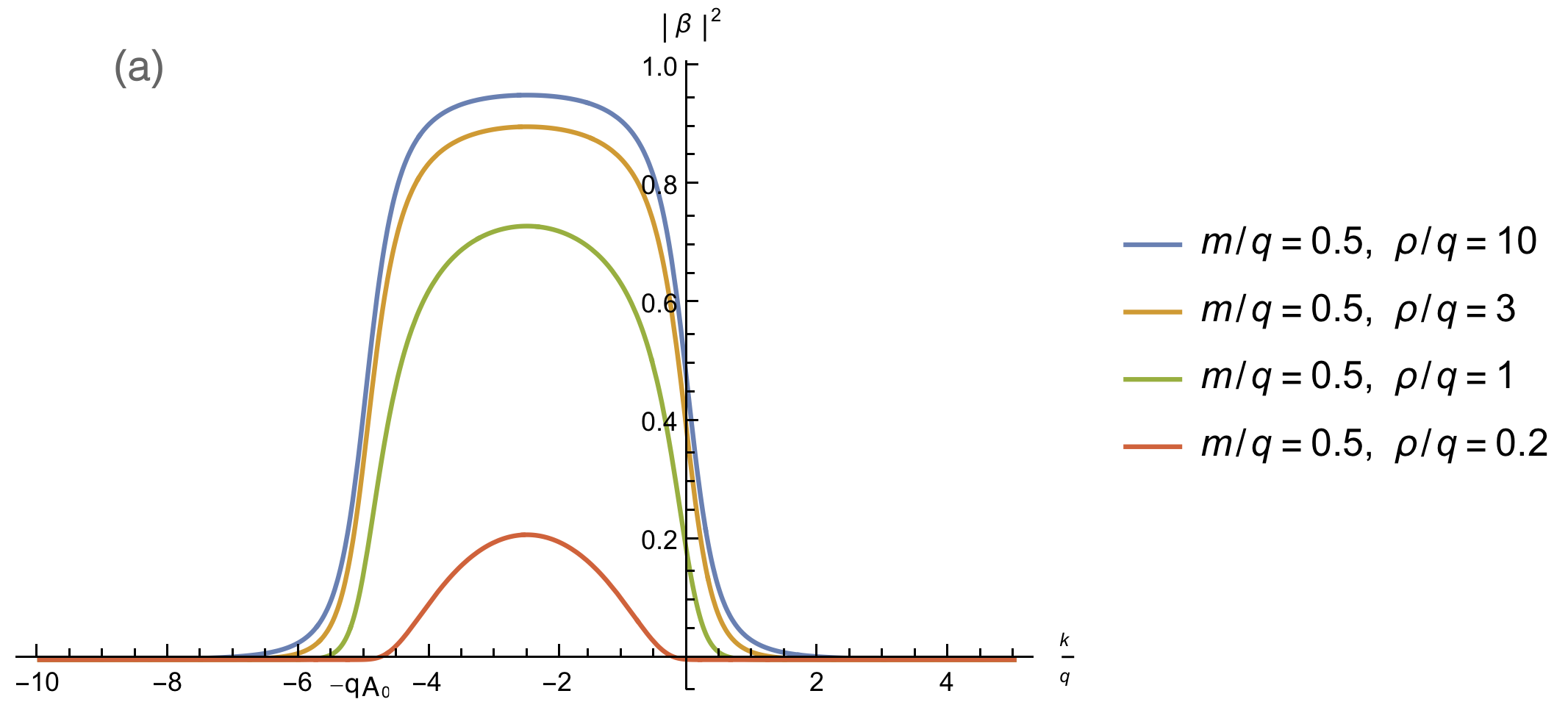}
\includegraphics[width=90mm]{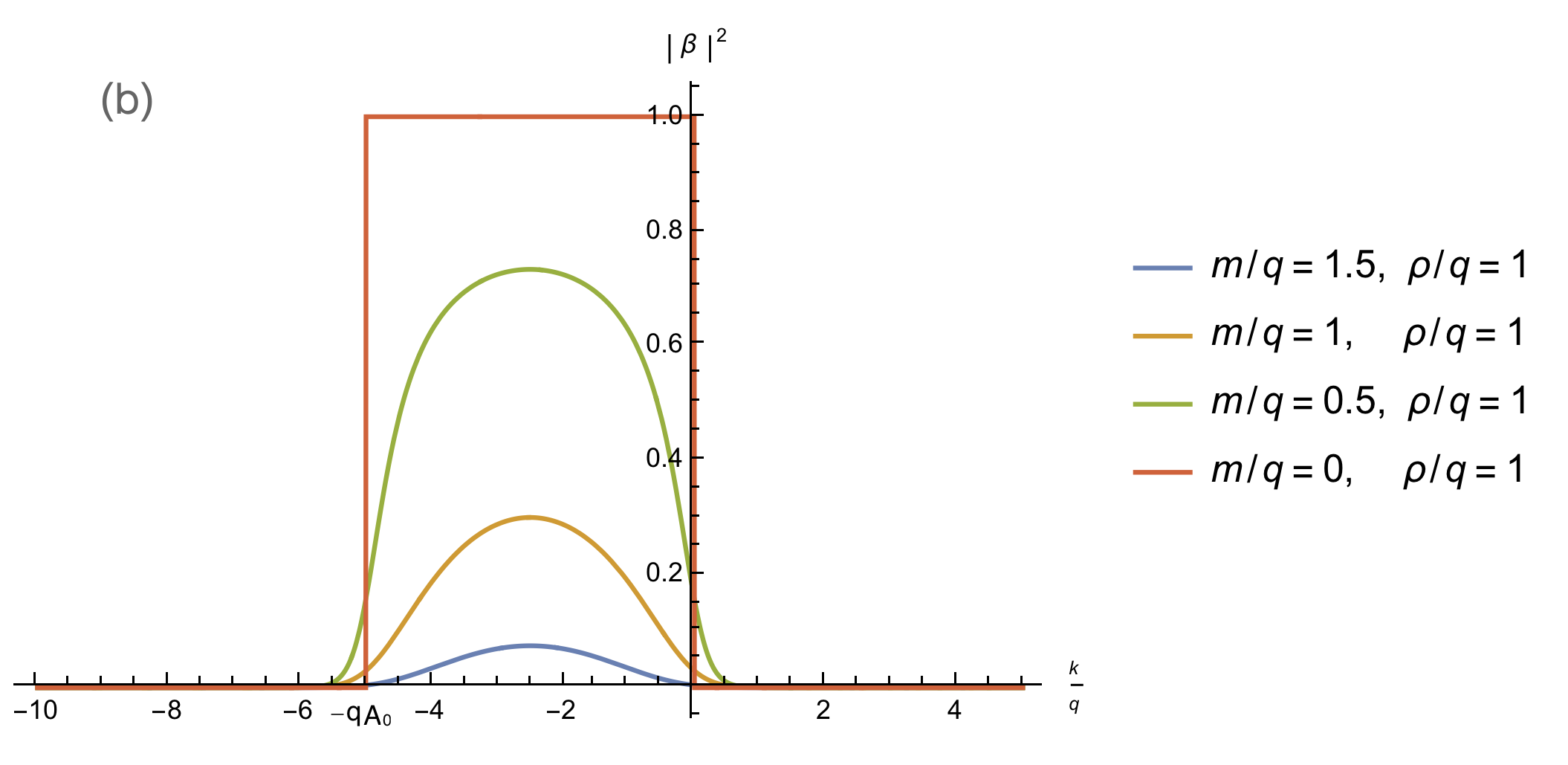}
\end{tabular}
\end{center}
\caption{\small{Momentum distribution of the created fermions with positive charge at late times $|\beta_k|^2$ by an electric pulse with $A_0=5$ and different values of $m/q$ and $\rho/q$. In figure (a) the mass is fixed, while in (b) the  parameter of adiabaticity is fixed. Note that the relative position of the of the curves $|\beta_k|^2$ with respect to the vertical axis is different  to the scalar plots because of the  opposite convention for the electric charge, as explained in the main text. }}
\label{betafermion}
\end{figure}

Let us consider now a two-dimensional charged Dirac field $\psi$ interacting with an homogeneous, time-dependent electric field. The corresponding Dirac equation  is
 \bea
 (i \gamma^{\mu}D_{\mu}-m)\psi=0\label{diraceq} \  ,
 \eea
 where $\gamma^{\mu}$ are the Dirac matrices satisfying the anticommutation relations $\{\gamma^{\mu},\gamma^{\nu}\}=2\eta^{\mu\nu}$ and $D_{\mu} \equiv \partial_{\mu} -i q A_{\mu}$.  [We follow here the convention that the electric charge of the fermion is  $-q$]. The electromagnetic field is assumed to be an external classical field, while $\psi$ is a quantized  field interacting with the classical electric background.
Assuming also that the electric field is described by the potential $A_{\mu}= (0, -A(t))$ in the appropriate gauge, the Dirac equation \eqref{diraceq} becomes 
  \bea
 \label{Dirac}\left(i \gamma^0 \partial_0 +\left(i\partial_x-q A\right)\gamma^1-m\right)\psi=0.
 \eea
From now on we will use  the Weyl representation (with $\gamma^5 \equiv \gamma^0\gamma^1$)
\bea
\gamma^0 = \scriptsize
\left( {\begin{array}{cc}
 0 & 1  \\
 1& 0  \\
 \end{array} } \right),\hspace{2cm} 
\gamma^1 = \scriptsize \left( {\begin{array}{cc}
 0 & 1  \\
 -1& 0  \\
 \end{array} } \right), \hspace{2cm} \gamma^5 = \scriptsize \left( {\begin{array}{cc}
 -1 & 0  \\
 0& 1 \\
 \end{array} } \right) \nonumber
 \ . \eea
\\
We expand the field in  momentum modes 
\bea
\label{spinorbd}\psi(t, x)=\int_{-\infty}^{\infty} dk \left[B_k u_k(t, x)+D^{\dagger}_k v_k(t, x)\right] \ , 
\eea
where the two independent and normalized spinor solutions are
\bea
 u_{k}(t, x)&=&\frac{e^{ikx}}{\sqrt{2\pi }} \scriptsize \left( {\begin{array}{c}
 h^{I}_k(t)   \\
 -h^{II}_k (t) \\
 \end{array} }\right) \\
  v_{k}(t, x)&=&\frac{e^{-ikx}}{\sqrt{2\pi }} \scriptsize \left( {\begin{array}{c}
 h^{II*}_{-k} (t)  \\
 h^{I*}_{-k}(t)  \\
 \end{array} } \right)
 \label{spinorde}
\ . \eea
$B_k$ and $D_k$ are the creation and annihilation operators which fulfill the usual anti-commutation relations. The field equation (\ref{Dirac}) is converted into 
\bea \label{system}
&&\dot{h}^{I}_k-i\left(k+qA\right)h^{I}_k-i m h^{II}_k=0\\ \label{system2}
&&\dot{h}^{II}_k+i\left(k+qA\right)h^{II}_k-i m h^{I}_k=0 \ , 
\eea
and we have assumed  the normalization condition  
 $|h_k^{I}|^2+|h_k^{II}|^2=1$. 
Let us consider, as in the scalar case, the electric pulse $A(t)=\frac12 A_0\left(\tanh(\rho t)+1\right) \label{potentialaf}$. With this input  the mode equations  \eqref{system} and \eqref{system2} can be  solved exactly in terms of hypergeometric functions
\bea \label{solutionh1}
h_k^{I}(t)&=& \sqrt{\frac{\omega_{in}-k}{2\omega_{in}}} \left(\frac{A(t)}{A_0}\right)^{-i\frac{\omega_{in}}{2\rho}}\left(1-\frac{A(t)}{A_0}\right)^{i\frac{\omega_{out}}{2\rho}} F\left(i\frac{\omega_-+qA_0/2}{\rho},1+i\frac{\omega_--qA_0/2}{\rho},1-i\frac{\omega_{in}}{\rho};\frac{A(t)}{A_0}\right) 
 \eea

 \bea \label{solutionh2}
 h_k^{II}(t)&=& -\sqrt{\frac{\omega_{in}+k}{2\omega_{in}}} \left(\frac{A(t)}{A_0}\right)^{-i\frac{\omega_{in}}{2\rho}}\left(1-\frac{A(t)}{A_0}\right)^{i\frac{\omega_{out}}{2\rho}} F\left(i\frac{\omega_--qA_0/2}{\rho},1+i\frac{\omega_-+qA_0/2}{\rho},1-i\frac{\omega_{in}}{\rho};\frac{A(t)}{A_0}\right)
 \eea
 where $\omega_{in}=\sqrt{k^2 + m^2}$, $\omega_{out}=\sqrt{(k+qA_0)^2 + m^2}$  and $\omega_\pm=\frac12 (\omega_{out} \pm \omega_{in})$. We have fixed the initial condition in order to recover the positive frequency solution for a free field at early times $t \to -\infty$ 
\bea
h_k^{I/II}(t)\sim  \pm \sqrt{\frac{\omega_{in} \mp k}{2 \omega_{in}}}e^{- i \omega_{in} t} \ .
\eea
At late times $t \to +\infty$ the modes can be written  as 
\bea \label{h1inf}
h_k^{I/II}(t)\sim  \pm \sqrt{\frac{\omega_{out} \mp (k+qA_0)}{2 \omega_{out}}}\alpha_ke^{- i \omega_{out} t} + \sqrt{\frac{\omega_{out} \pm (k+qA_0)}{2 \omega_{out}}}\beta_ke^{i \omega_{out} t} \ .
\eea
$\alpha_k$ and $\beta_k$ are the Bogoliubov coefficients  satisfying  the relation $|\alpha_k|^2+|\beta_k|^2=1$. These coefficients relate the early time creation and annihilation operators ${B_k,D_k}$ with the late time operators ${b_k,d_k}$ as follows
\bea
b_k&=&\alpha_k B_k+\beta_{k}^{*}D_{-k}^{\dagger}\\
d_{k}&=&\alpha_{-k}D_{k}-\beta_{-k}^{*}B_{-k}^{\dagger} \ .
\eea
The density of created quanta is given by $N_k= \bra{0}b_k^{\dagger}b_k\ket{0}+\bra{0}d_k^{\dagger}d_k\ket{0}\equiv n_k + \bar{n}_k$, where $n_k= |\beta_k|^2$ and $\bar n_k= |\beta_{-k}|^2$. Therefore, the particle number is also \bea
\langle N \rangle=\frac{1}{2 \pi}\int_{-\infty}^{\infty}dk~ N_k=\frac{1}{2 \pi}\int_{-\infty}^{\infty}dk\left(|\beta_{k}|^2+|\beta_{-k}|^2\right) \ . 
\eea
The matching of (\ref{solutionh1}-\ref{solutionh2})  with \eqref{h1inf} at late times determines the Bogoliubov coefficients. For the beta coefficients we get
\bea
\beta_k=\sqrt{\frac{\omega_{out}}{\omega_{in}}\frac{\omega_{in}-k}{\omega_{out}+k+qA_0}}   \frac{\Gamma(1-i\frac{\omega_{in}}{\rho})\Gamma(-i \frac{\omega_{out}}{\rho})}{\Gamma(1 + i\frac{\omega_{-}+ qA_0/2}{\rho})\Gamma(1 +i\frac{\omega_{-}- qA_0/2}{\rho})} \ . 
\eea
And after simplifying, we obtain
 \bea |\beta_{k}|^2=\frac{\cosh{(2\pi \frac{\omega_{-}}{\rho})}-\cosh{(\pi \frac{qA_0}{\rho})}}{2\sinh{(\pi \frac{\omega_{in}}{\rho})}\sinh{(\pi \frac{\omega_{out}}{\rho})}} \label{betaqedf}\ .\eea 
Some representations of this expression are shown in Fig.\ref{betafermion}. As in the scalar case, the number of particles decreases as $\rho \to 0$ and increases as $m\to0$. For fermions, the relation $|\alpha_k|^2+|\beta_k|^2=1$ implies that $|\beta_k|^2\leq 1$ for any value of $k$, according to  Pauli's exclusion principle.\\

In the massless case, irrespective of the value of $\rho$, one obtains (see Fig. \ref{betafermion}(b))
\bea
\lim_{m\to 0} |\beta_k|^2=1 
\eea
for $k\in(0,qA_0) $, and hence
\bea
N_k= \left\{ \begin{array}{cc}
   0 & \text{ for } k\notin(-qA_0,qA_0) \  \ \ \ \\
   1 & \text{ for } k\in(-qA_0,qA_0)   \ \ \ .  \\

 \end{array} \right.
 \eea 
The total density of created quanta is
\bea
\langle N \rangle=\frac{1}{2\pi}\int_{-|q A_0|}^{|qA_0|} dk~ N_k=\frac{|qA_0|}{\pi} \label{particleprod}.
\eea 
Note that the same result is obtained by performing the adiabatic limit $\rho \to 0$ in the scalar case. In contrast, this result is valid for any value of $\rho$, which means that the number of created massless fermions does not depend on  the history of $A(t)$, but only on its final value. 
 This non-vanishing result of the particle number implies again the breaking of the adiabatic invariance.\\

\begin{figure}[htbp]
\begin{center}
\includegraphics[width=90mm]{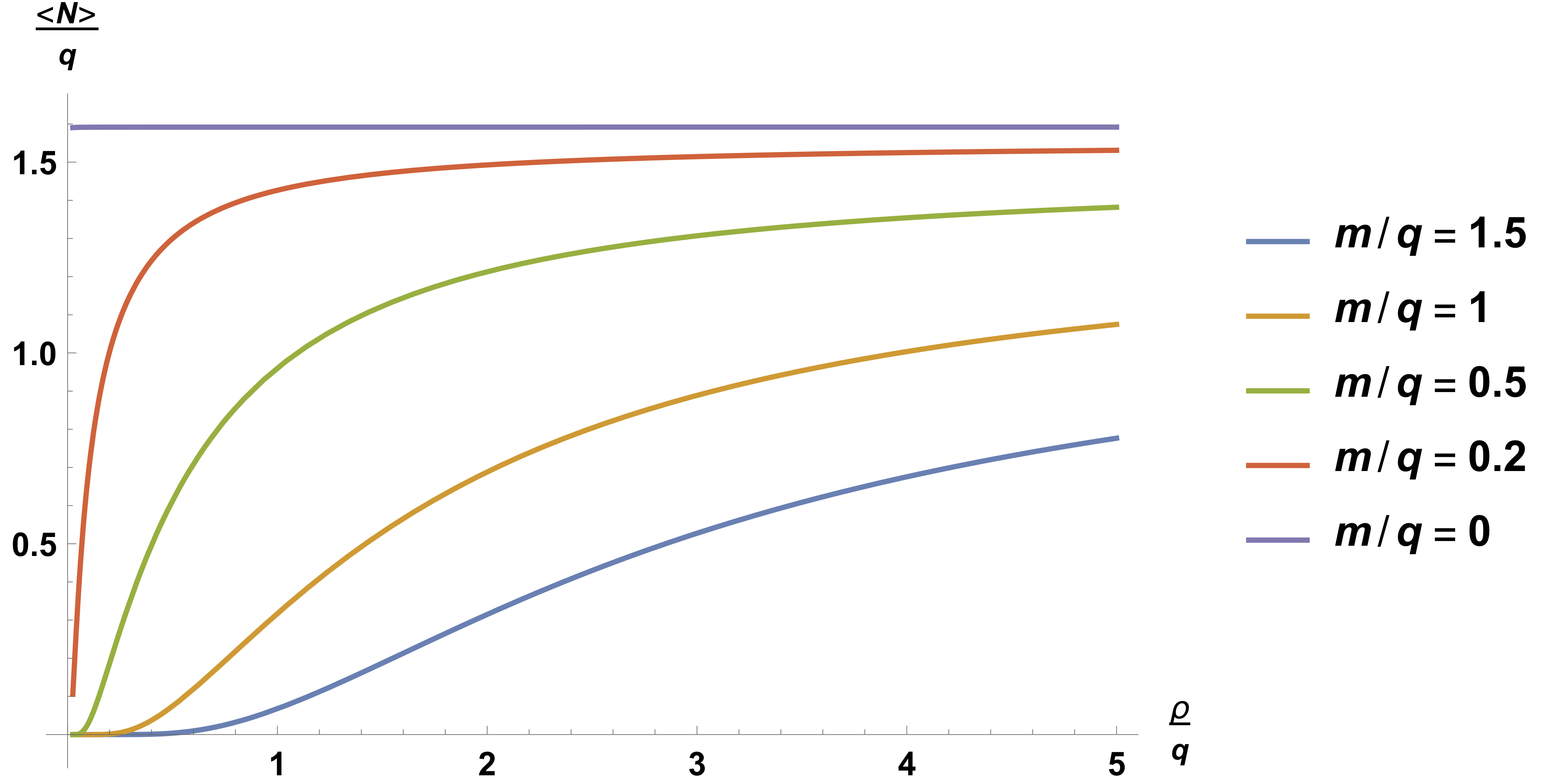}
\end{center}
\caption{\small{Number of late-time created fermions  as a function  of the dimensionless adiabaticity  parameter  $\rho/q$, for $A_0=5$ and for different values of the mass.}}
\label{Nfermion}
\end{figure}

For massive fermions and in the limit $\rho \to 0$, expression \eqref{betaqedf} behaves essentially as 
\be
|\beta_k|^2  \sim 
e^{-\frac{\pi}{\rho} \delta} \ , \label{betalimitf}
\ee
where $\delta=2\omega_+-|qA_0|$. For $m\neq0$, the former has a minimum at $k=-\frac{qA_0}{2}$, with a value $\delta_{min}=\sqrt{(qA_0)^2+4m^2}-|qA_0|>0$. Hence, $\delta>0$ and $|\beta_k|^2 \to 0$, as we can see in Fig. \ref{betafermion}(a). Therefore we can conclude that the particle number  is an adiabatic invariant for massive fermions, as in the scalar case. To visualize this behavior, we have depicted in Fig. \ref{Nfermion} the dependence of the total density of created particles on the parameter $\rho$. We can also observe that the density of quanta in the massless case does not vanish and, in contrast to the scalar case, it remains constant, according to the above calculations.

\subsection{Electric current}

Using the renormalization method described in \cite{FN, BFNV, BNP} for a  Dirac field interacting with an homogeneous time-dependent electric field, the vacuum expectation value of the electric current  $j^\mu = -q\bar \psi \gamma^\mu  \psi$ is given by 
	
	\bea
	\langle j^x \rangle_{ren}=\frac{q}{2\pi } \int dk \left( |h_k^{II}|^2-|h_k^{I}|^2-\frac{k}{\omega}-\frac{q m^2} {\omega^3}A\right)\label{corriente1} \ .
	\eea	
To study the explicit dependence of the electric current $\langle j^x \rangle$ with the mass, we can compute their time derivative 

\be \partial_t\langle j^x \rangle_{ren}   = \frac{2qm}{\pi} \left(\int  Im( h_k^{II}h_k^{I*})dk\right) -\frac{q^2} {\pi  }\dot A \label{Dj}. \ee
It is immediate to see that in the massless limit the first term vanishes, and the equation below can be easily integrated. With $A(-\infty)=0$ as initial condition one obtains
\bea
\label{current} \langle j^x \rangle_{ren}= -\frac{q^2A(t)}{\pi}  \ .
\eea

Note again that at each instant $t$, the value of the electric current depends only on the value of the potential vector $A(t)$ at $t$, and not on its history.
This contrasts with the behavior of the created current for a massive field, since $h^{I}_k$ and $h^{II}_k$ depend on the particular shape of the electric pulse. In the latter case, and assuming the electric pulse configuration given in \eqref{potentiala}, it can be proven by using the Bogoliubov coefficient \eqref{betaqedf} and repeating the same calculations done in Section IIIA, that the electric current vanishes in the adiabatic limit. The analysis of the renormalized energy density can be carried out analogously, and it leads to a similar physical conclusion. \\

As a final comment, and for completeness,  we remark that in the massless case, one can directly solve the semiclassical Maxwell equations for the electric field $\dot E = - \langle j^x\rangle_{ren}$. Assuming, for instance, the  initial conditions $A(0)=0$ and $\dot{A}(0)=-E_0$ the previous equation can be easily integrated, with solution
$ E(t)=E_0\cos (\frac{\mid q\mid}{\sqrt{\pi}} t )$. We find harmonic oscillations with frequency $\frac{|q|}{\sqrt{\pi} }$. This  is consistent with the well-known fact that radiative corrections to the Schwinger model 
induce a mass for the ``photon'', with a value $m_{\gamma}^2 = q^2/\pi$ \cite{Schwinger62, qftbook}.


\subsection{Relation with the axial anomaly} \label{relation}

We have found that the expected adiabatic invariance of the particle number observable fails for a massless Dirac field. This is accompanied with a non-vanishing electric current, even in the adiabatic limit, as can be read from (\ref{current}). Furthermore, this result brings about a creation of chirality as a consequence of the fact that, in two-dimensions, the axial current $j^\mu_5 = \bar \psi \gamma^\mu \gamma^5 \psi$ is related by duality  to the electric current $ \langle j_\mu \rangle_{ren} = q\epsilon_{\mu\nu} \langle j^{\nu}_{5} \rangle_{ren}$. Hence, the result (\ref{current}) implies the axial anomaly \cite{Bertlmann} 
\be  \partial_\mu \langle j^\mu_5 \rangle_{ren} = -\frac{q}{2\pi} \epsilon^{\mu\nu} F_{\mu\nu} \ . \ee
 
In fact, one can also interpret the breaking of the adiabatic invariance as a natural and necessary consequence required by the axial anomaly. We remark that the loss of the adiabatic invariance of the particle number for a scalar field in two-dimensions, which coincides quantitatively with the result for fermions, can also be naturally interpreted in the language  of  anomalies. In two-dimensions,  a massless scalar field inherits a classical chiral-type symmetry, in the sense that the classical wave equation splits into two disconnected sectors: right and left-moving degrees of freedom, as the fermionic two-dimensional field. The corresponding right and left electric currents are, in the adiabatic limit, separately conserved in the classical theory. However, in the quantum theory these currents also cease to be conserved. The creation of  right and left electric currents in the quantum theory is exactly the same for massless scalar and Dirac fields in the adiabatic limit, as can be easily observed from (\ref{current}) and (\ref{jbetam0}).


\section{ Generalization of previous results to 4D}

In the previous sections we have shown that the particle number operator is not an adiabatic invariant for two-dimensional massless  fields. 
Here, we extend our analysis to four dimensions for both scalar and fermionic fields.  We  briefly study whether the breaking of the adiabatic invariance could also happen in electric and magnetic backgrounds. 

\subsection{Scalar field}

Consider  now a charged scalar field obeying the wave equation $(D_\mu D^\mu + m^2) \phi =0$, where we assume an homogeneous electric pulse defined by the vector potential $A_\mu= ( 0, 0, 0, -A(t))$ with $A(t)$ given again by \eqref{potentiala}.
The Fourier expansion of the quantized field is  
\be \phi(t,\vec{x})= \frac{1}{\sqrt{2(2 \pi)^3}}\int d^3 k [A_{\vec k}e^{i \vec k \vec x}h_{\vec k}(t)+B_{\vec k}^{\dagger}e^{-i \vec k \vec x}h^*_{-\vec k}(t) ] \ . \label{phisolution2} \ee
The mode functions $h_{\vec  k}(t)$ satisfy the normalization condition $ h_{\vec  k}\dot h_{\vec k}^* - h_{ \vec k}^*\dot h_{ \vec k} = 2i$ and their time evolution is given by

\be
\ddot{h}_{\vec k}(t)+\left(m^2+ k_1^2 + k_2^2 + (k_3-q A(t))^2\right)h_{\vec k}(t)=0  \ . \label{equhk2} \ee

  This equation is very similar to the one found in the two-dimensional case \eqref{equhk}.    It allows us to partially reduce the four-dimensional problem to a two-dimensional one, by introducing an effective mass  $m^2_{eff}= m^2+ k_1^2 + k_2^2$. Therefore, the beta coefficients can be obtained from Eq. \eqref{betaqed} replacing $k$ by $k_3$ and $m$ by $m_{eff}$. \\

  According to our previous results for scalar fields, only for $m=0$ and $k_1=0=k_2$ one can have a non-vanishing  $|\beta_{k}|^2$  in the adiabatic limit. However, since $k_1$ and $k_2$ are continuous quantum numbers characterizing the modes, the amount of created particles $\langle N \rangle \sim \int d^3k(|\beta_{k}|^2+|\beta_{-k}|^2)$ is diluted into the infinite-volume of the unbounded three-dimensional space. Therefore, the total number density of produced particles turns out to be an adiabatic invariant. \\
  
  This result cannot be altered by the introduction of a magnetic field. Adding a constant magnetic field $\vec B$ in the $z$-direction and choosing  $A_\mu= (0, 0,- Bx^1, -A(t))$, the Fourier expansion for the scalar field is
  \bea \phi(t,\vec{x})= \frac{1}{\sqrt{2(2 \pi)^2}}\sum_n \int \int dk_2dk_3 &[&A_{n,k_2,k_3}e^{i(k_2x^2+k_3x^3)} \Phi_{n,k_2}(x^1) h_{n,  k_3}(t)\nonumber  \\ &+&B_{n,k_2,k_3}^{\dagger}e^{-i(k_2x^2+k_3x^3)} \Phi_{n,-k_2}(x^1) h^{*}_{n,-k_3}(t) \,] \ , \label{phisolution2} \eea
where
\be \label{hermite} \Phi_{n,k_2}(x^1) =\left(\frac{qB}{\pi}\right)^{1/4}\frac{1}{2^{\frac{n}{2}}\sqrt{n!}} e^{-\xi^2/2} H_n(\xi)  \ , \ee
 $\xi= \sqrt{qB}(x^1-k_2/qB)$, and  $H_n(\xi)$ are the Hermite polynomials with $n=0, 1, 2, ...$ \ . For simplicity, and without loss of generality, we have assumed $qB>0$. The time evolution is given by
\be \ddot h_{n, k_3} + \left(m^2+ (2n+1)qB+(k_3-q A(t))^2\right)h_{k}(t)=0 \ . \ee
From the two-dimensional viewpoint, the effective value of the mass, given now by  $m^2_{eff}=m^2+ (2n+1)qB $, is a positive quantity, even for $m=0$ and $n=0$. Using again the result of  Section III we can similarly conclude that the particle number, defined now as 
\be \label{pNB}
\langle N \rangle=\frac{qB}{4 \pi^2}\sum_{n=0}^{\infty}\int_{-\infty}^{\infty} dk_3~ N_{n,k_3}=\frac{qB}{4 \pi^2}\sum_{n=0}^{\infty}\int_{-\infty}^{\infty}dk_3\left(|\beta_{n,k_3}|^2+|\beta_{n,-k_3}|^2\right),
\ee
is also an adiabatic invariant for a  scalar field in four dimensions, regardless  of the value of the mass, given that $|\beta_{n,k_3}|^2\to 0$ . This is in sharp contrast with the result obtained for a massless scalar field in two dimensions. Note that for a scalar field in four dimensions there is no analog of the axial anomaly.

\subsection{Dirac field}

We can repeat the analysis for Dirac fermions. For massive fermions adiabatic invariance is preserved. Therefore we will focus on the massless case. In the latter, one can split the Dirac spinor  in two independent chiral parts $\psi=\scriptsize \left( {\begin{array}{c}
 \psi_L  \\
 \psi_R \\
 \end{array} }\right)$. For the left sector the Weyl equation reads $\partial_0\psi_L -\vec{\sigma} \vec{D} \psi_L=0$. Considering an homogeneous electric pulse with vector potential $A_\mu= ( 0, 0, 0, -A(t))$ given by \eqref{potentiala}, the Fourier expansion of the quantized field is  
\bea
\label{spinorbd}\psi_L(t, \vec{x})=\int d^3k \left[B_{\vec k} u_{\vec k}(t, \vec{x})+D^{\dagger}_{\vec k} v_{\vec k}(t, \vec{x})\right] \ . 
\eea

 The two independent and normalized spinor solutions can be expressed as:

\bea
 u_{\vec k}(t, \vec{x})&=&\frac{e^{i\vec k \vec x}}{(2\pi)^{3/2} \,k_\bot} \scriptsize \left( {\begin{array}{c}
 (k_1 - i k_2)\, h^{I}_{\vec k}(t)  \\
 k_\bot h^{II}_{\vec k} (t) \\
 \end{array} }\right) \\
  v_{\vec k}(t, \vec{x})&=&\frac{e^{-i\vec k\vec x}}{(2\pi)^{3/2} \,k_\bot} \scriptsize \left( {\begin{array}{c}
 (k_1 - i k_2)\, h^{II*}_{-\vec k} (t) \\
  k_\bot h^{I*}_{-\vec k}(t)\\
 \end{array} } \right)
 \label{spinorde}
\ , \eea
where $k_\bot=\sqrt{k_1^2+k_2^2}$. The equations for the modes are
\bea \label{systemt}
&&\dot{h}_{\vec k}^{I}-i\left(k_3+qA\right)h_{\vec k}^{I}-i\,k_\bot\,h_{\vec k}^{II}=0\\ \nonumber
&&\dot{h}_{\vec k}^{II}+i\left(k_3+qA\right)h_{\vec k}^{II}-i\,k_\bot\,h_{\vec k}^{I}=0 \ . 
\eea
  These equations are similar to the ones found in the two-dimensional case \eqref{system} and \eqref{system2}, with an effective mass  $m_{eff}=  k_\bot$. Hence, the beta coefficients are given by Eq. \eqref{betaqedf} with the obvious replacements.   
As in the scalar case, only for $k_1=k_2=0$ one can have a non-vanishing beta coefficient in the adiabatic limit, therefore the amount of created particles is diluted and the total number density of produced particles is an adiabatic invariant. However, this is no longer true in the presence of a magnetic field. Adding a constant magnetic field $\vec B$ in the $z$-direction and choosing  $A_\mu= (0, 0, Bx^1, -A(t))$, the generic form of the modes for a massless field is
\bea \label{ansatz1}
u_{n,k_2,k_3}(t, \vec{x})&=&\frac{e^{i(k_2x^2+k_3x^{3})}}{2 \pi} \scriptsize \left( {\begin{array}{c}
 h_{n,k_3}^{I}(t) \Phi_{n,k_2}(x^1)  \\
 -i h_{n,k_3}^{II}(t) \Phi_{n-1,k_2}(x^1)  \\
 \end{array} }\right) \\
 v_{n,k_2,k_3}(t, \vec{x})&=&\frac{e^{-i(k_2x^2+k_3x^{3})}}{2 \pi} \scriptsize \left( {\begin{array}{c}
 h_{n,-k_3}^{II*}(t) \Phi_{n,-k_2}(x^1)   \\
 i h_{n,-k_3}^{I*}(t) \Phi_{n-1,-k_2}(x^1)  \\
 \end{array} }\right)
\ , \eea
where $\Phi_{n,k_2}$ is defined as in the scalar case \eqref{hermite}. The time evolution of the modes is given by
\bea \label{systemt2}
&&\dot{h}_{n,k_3}^{I}-i\left(k_3+qA\right)h_{n,k_3}^{I}-i \sqrt{2nqB} h_{n,k_3}^{II}=0\\ \nonumber
&&\dot{h}_{n,k_3}^{II}+i\left(k_3+qA\right)h_{n,k_3}^{II}-i \sqrt{2nqB} h_{n,k_3}^{I}=0 \ .
\eea
In this case, we can  identify the effective mass as $m^2_{eff}= 2nqB$, which vanishes at $n=0$. 
Therefore, in the adiabatic limit, the beta coefficients $|\beta_{n,k_3}|^2 $ (we recall they can also be obtained from the two-dimensional analog \eqref{betaqedf}) vanish for any value of $n$ except for $n=0$. Since $\langle N \rangle \sim \sum_{n} \int dk_3(|\beta_{n,k_3}|^2+|\beta_{n,-k_3}|^2)$ , the particle number tends to a non-zero value because the discrete state $n=0$ survives after summation. This contrasts with the previous case in which the mode $k_1=k_2=0$  was diluted after integration. Hence, the particle number $\langle N \rangle $ is no longer adiabatic invariant. \\

This result is also linked to  the axial anomaly, as happens in two dimensions. Note that in four dimensions the anomaly is only non-zero when both electric and magnetic fields are present.
As in the two-dimensional case, the adiabatic anomaly must be reflected in the electric current  $\langle j^z \rangle= -q\langle \bar \psi \gamma^3\psi \rangle$ and also in the chiral charge density $\langle j^0_5 \rangle= \langle \bar \psi \gamma^0\gamma^5\psi \rangle$. Repeating the previous analysis for the right part $\psi_R$ and computing the formal vacuum expectation value $\langle j^z\rangle$ one finds

\bea \label{j3}
\langle j^{z}\rangle=  \frac{q^2B}{4 \pi^2}\int_{-\infty}^{\infty}dk_3(|h_{0, k_3}^{II}|^2-|h_{0,k_3}^{I}|^2)+\frac{q^2B}{2 \pi^2}\sum_{n=1}^{\infty}  \int_{-\infty}^{\infty}dk_3\left(|h^{II}_{n,k_3}|^2-|h^{I}_{n,k_3}|^2\right) .
\eea
From this result one can easily see the special role of the $n=0$ modes, which are the only ones contributing to the 
breaking of the adiabatic invariance. Although in the most general case all the modes contribute to the electric current, in the adiabatic limit the contribution of the modes with $n>0$, for which $m_{eff}\neq 0$, will  vanish, as happens in the two-dimensional case. This gives us  a lower bound for the current. 
On the other hand, by looking at the chiral charge
\bea \label{j5}
\langle j^{0}_5\rangle=  \frac{qB}{4 \pi^2}\int_{-\infty}^{\infty}dk_3(|h_{0, k_3}^{I}|^2-|h_{0,k_3}^{II}|^2)\ , 
\eea
one realizes  that only the mode with $n=0$ creates chirality, even in a non-adiabatic regime. Furthermore, it is immediate to see that the lower bound of the electric current is given by $\langle j^{z}\rangle_{min}=-q\langle j^{0}_5\rangle$.\\

Note that (\ref{j5}) can be renormalized using the adiabatic prescription in two dimensions  (see Eq. (\ref{corriente1})) and the result is compatible with the axial anomaly $   \langle j^{0}_5\rangle_{ren} (t)= -\frac{q^2}{2\pi^2} \int_{-\infty}^t dt' \vec E (t') \vec B$. It can be easily argued that  a similar result can also be obtained for  a time-dependent magnetic field. \\

\section{Conclusions}

We have reexamined the adiabatic invariance of the particle number operator  of quantized fields in two dimensions coupled to a background electric field with bounded vector potential. We have pointed out that, for massless fields, the expected adiabatic invariance fails. This fact is accompanied by the emergence of the axial anomaly in two dimensions. In other words, the breaking of the adiabatic invariance (pair creation even in the limit $\rho \to 0$) is required to keep physical consistency with the axial anomaly. We have also shown that the breaking of the adiabatic invariance  is also reproduced for a massless Dirac field in four dimensions,  but requiring  the presence of electric and magnetic fields,  showing up again  a deep connection with  the axial anomaly. \\

{\it Acknowledgments.--}  We thank I. Agullo and  A. del Rio for very useful discussions. This work was supported by  Grants.  No.\  FIS2017-84440-C2-1-P, No. \ FIS2017-91161-EXP, No. \ SEJI/2017/042 (Generalitat Valenciana), 
No. \  SEV-2014-0398. P. B. is supported by a Ph.D. fellowship, Grant No.  FPU17/03712. S. P.  is supported by a Ph.D. fellowship, Grant No. FPU16/05287. 
A. F. is supported by the Severo Ochoa Ph.D. fellowship, Grant No. SEV-2014-0398-16-1, and the European Social Fund.


\begin{thebibliography}{99}



\bibitem{parker66} L. Parker, {\it The creation of particles in an expanding universe}, Ph.D. thesis, Harvard University (1966).  Dissexpress.umi.com, Publication
Number  7331244; {\it Phys.~Rev.~Lett.} {\bf 21}, 562 (1968); {\it Phys.~Rev.~D} {\bf 183}, 1057 (1969); {\it Phys.~Rev.~D} {\bf 3}, 346 (1971).

\bibitem{parker-toms}L.~Parker and D.~J.~Toms, {\it Quantum Field Theory in Curved Spacetime: Quantized Fields
and Gravity}, Cambridge University Press, Cambridge, England (2009).
\bibitem{birrell-davies} N.~D.~Birrell  and P.~C.~W.~Davies, {\it Quantum Fields in Curved Space}, Cambridge University Press, Cambridge, England (1982).

\bibitem{parker2012} L. Parker, {\it J.\ Phys.\ A}  {\bf 45},  374023 (2012).

\bibitem{Pittrich-Gies} W. Pittrich and H. Gies, {\it Probing the Quantum Vacuum}, Springer, Heidelberg (2000).
\bibitem{ELI}The extreme light infrastructure (ELI) project: www.extreme-light-infrastructure.eu/



\bibitem{Schwinger51} J. Schwinger, {\it Phys. Rev. } {\bf 82}, 664 (1951).


\bibitem{reheating}  L. A.  Kofman, A. D. Lindle, and A. A. Starobinsky, {\it Phys. Rev. Lett.} {\bf 73}, 3195 (1994); {\it Phys. Rev. Lett.}  {\bf 76}, 1011 (1996).


\bibitem{Mueller} N. Mueller, F. Hebenstreit, and J. Berges {\it Phys. Rev. Lett.} {\bf 117}, 061601 (2016); S. P. Kim and H. K. Lee {\it Phys. Rev. D} {\bf 76}, 125002 (2007).

\bibitem{Dunne} G. V. Dunne, {\it Int. J. Mod. Phys. A} {\bf 27}, 1260004 (2012); {\it  Eur. Phys. J. D} {\bf 55}, 327 (2009); B. S. Xie, Z. L. Lie, and S. Tang, {\it Matter Radiat. Extremes}, {\bf 2}, 225 (2017).

\bibitem{R} R. Ruffini, G. Vereshchagin and S. Xue, {\it Phys. Rep.} {\bf 487}, 1 (2010).

\bibitem{qftbook} M.E. Peskin and D. V. Schroeder, {\it An Introduction to Quantum Field Theory}, Addison-Wesley, Reading, MA, USA (1995).

\bibitem{Israel} W. Israel, {\it Phys. Rev. Lett.} {\bf 57}, 397 (1986).

\bibitem{Hawking} S. W. Hawking, {\it Black Holes aren't Black}, Gravity Research Foundation (1974), essay [www.gravityresearchfoundation
738 .org/year\#1974]; {\it Commun. Math. Phys.} {\bf 43}, 199 (1975).

\bibitem{FabbriNavarro} A. Fabbri and J. Navarro-Salas, {\it Modeling Black Hole Evaporation}, ICP-World Scientific, London (2005).


\bibitem{Sauter} F. Sauter, {\it Z. Phys.} {\bf 69}, 742 (1931).

\bibitem{FN} A. Ferreiro and J. Navarro-Salas, {\it Phys. Rev. D} {\bf 97}, 125012   (2018).

\bibitem{mathbook} M. Abramowitz, and I. A. Stegun, {\it Handbook of Mathematical Functions: With Formulas, Graphs and Mathematical Tables}, Dover, New York (1972).

\bibitem{FNP} A. Ferreiro, J. Navarro-Salas and S. Pla, {\it Phys. Rev. D} {\bf 98}, 045015 (2018).



\bibitem{parker-fulling} L.~Parker and S.~A.~Fulling, {\it Phys.~Rev.~D} {\bf 9}, 341 (1974); S. A. Fulling and L. Parker, {\it Ann.~Phys.} (N.Y.) {\bf 87}, 176 (1974); P.~R.~Anderson and L.~Parker, {\it Phys.~Rev.~D} {\bf 36}, 2963 (1987);  I.~Agullo, J.~Navarro-Salas, G.~J.~Olmo and  L.~Parker, {\it Phys.~Rev.~Lett.} {\bf 103}, 061301 (2009);
{\it Phys.~Rev.~D} {\bf 81}, 043514, (2010); A. Landete, J. Navarro-Salas and F. Torrenti, {\it Phys. Rev. D} {\bf 89} 044030 (2014); S. Ghosh, {\it Phys. Rev. D} {\bf 91}, 124075 (2015); I. Agullo, W. Nelson and A. Ashtekar {\it Phys. Rev. D} {\bf 91},  064051 (2015); A.~del Rio and  J.~Navarro-Salas, {\it Phys.~Rev.~D} {\bf 91}, 064031 (2015); S. Ghosh, {\it Eur. Phys. J. C} {\bf 79}, 239 (2019);
A. Ferreiro and J. Navarro-Salas, {\it Phys. Lett. B} {\bf 792}, 81 (2019). 
\bibitem{BFNV}J. F. Barbero G., A. Ferreiro, J. Navarro-Salas and E. J. S. Villase\~nor, {\it Phys. Rev. D} {\bf 98}, 025016 (2018).  
\bibitem{BNP} P. Beltr\'an-Palau, J. Navarro-Salas and S. Pla,  {\it Phys. Rev. D} {\bf 99} 105008 (2019).




























\bibitem{Schwinger62} J. Schwinger, {\it Phys. Rev.} {\bf 128}, 2425 (1962).

\bibitem{Bertlmann} R. A. Bertlmann, {\it Anomalies in Quantum Field Theory}, Oxford University Press, Oxford (2000).




















%
%
%
%























\end{thebibliography}
\end{document}